%% file: arXiv.tex
\documentclass[10pt,twocolumn,letterpaper]{article}

\usepackage[pagenumbers]{cvpr} 

\usepackage{graphicx}
\usepackage{amsmath}
\usepackage{amssymb}
\usepackage{booktabs}
\usepackage{tabularx}
\usepackage{multirow}
\newcolumntype{Y}{>{\centering\arraybackslash}X}
\usepackage{xcolor}
\newcommand{\greencheck}{{\color{green}\checkmark}}
\usepackage{pifont}
\usepackage[bottom]{footmisc}
\usepackage{algorithm}
\usepackage{algpseudocode}

\newcommand{\xmark}{\ding{55}}
\newcommand{\redcross}{{\color{red}\xmark}}

\usepackage[pagebackref,breaklinks,colorlinks]{hyperref}

\usepackage[capitalize]{cleveref}
\crefname{section}{Sec.}{Secs.}
\Crefname{section}{Section}{Sections}
\Crefname{table}{Table}{Tables}
\crefname{table}{Tab.}{Tabs.}

\def\cvprPaperID{337} 
\def\confName{CVPR}
\def\confYear{2022}

\newcommand{\OurModel}{E-CIR}

\begin{document}
\title{\OurModel: Event-Enhanced Continuous Intensity Recovery}

\author{Chen Song \quad Qixing Huang \quad Chandrajit Bajaj\\
The University of Texas at Austin\\
{\tt\small \{song, bajaj, huangqx\}@cs.utexas.edu}
}
\maketitle

\input{00-Abstract}
\input{01-Introduction}

\input{02-RelatedWork}
\input{03-Preliminary}
\input{04-Method}
\input{05-Evaluation}

\input{06-Limitations}
\input{07-Conclusion}
\input{08-Acknowledgement}

\section{Supplementary Material}
The source code of \OurModel, scripts and step-by-step instructions to converting the REDS~\cite{Nah_2019_CVPR_Workshops_REDS} dataset to the event format are available at \href{https://github.com/chensong1995/E-CIR}{https://github.com/chensong1995/E-CIR}. Additional animations can be viewed at \href{http://songc.me/ecir/visualization.html}{http://songc.me/ecir/visualization.html}. The static version of these additional results are presented in the remaining pages of this .pdf file. 

\section{Potential Negative Societal Impact}
In our experiments, we train \OurModel~using three Tesla V100-SXM2-32GB GPUs for approximately 100 hours. The total emission is estimated to be 38.88 kgCO$_2$eq, equivalent to 157.2 km driven by an average car. This estimation is conducted using the \href{https://mlco2.github.io/impact#compute}{Machine Learning Impact calculator} presented in \cite{lacoste2019quantifying}. To mitigate repetitive labor and negative environmental impact in future research, we have released our open-source implementation together with trained network weights.

\clearpage

\begin{figure*}
    \centering
    \includegraphics[width=\textwidth]{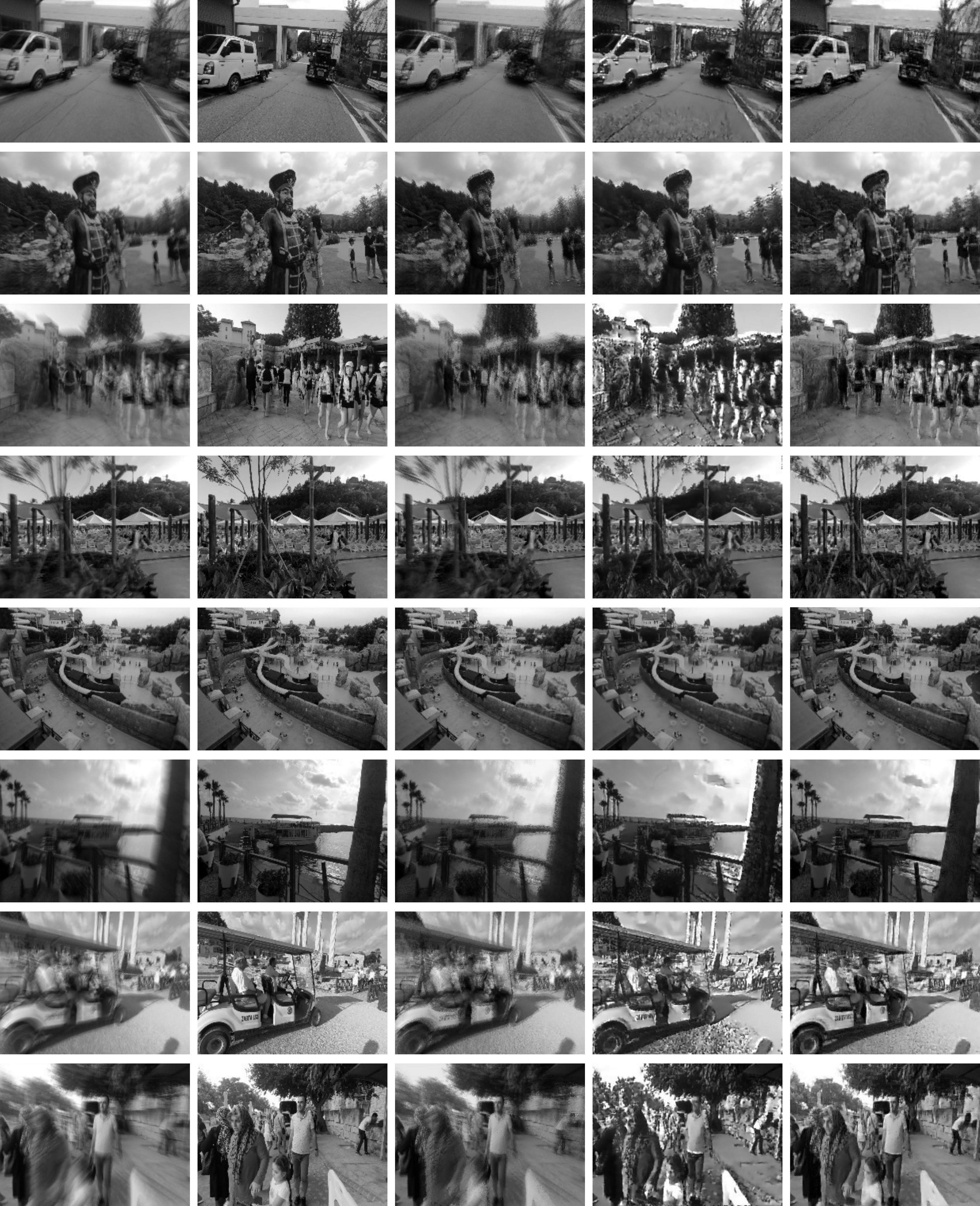}
    \begin{tabularx}{\textwidth}{Y Y Y Y Y}
        Input & Ground Truth & EDI~\cite{pan2019bringing} & eSL-Net~\cite{wang2020event} & Ours 
    \end{tabularx}
    \vspace{-0.2in}
    \caption{Qualitative visualization on the REDS~\cite{Nah_2019_CVPR_Workshops_REDS} dataset.}
    \label{fig:reds_2}
\end{figure*}

\begin{figure*}
    \centering
    \includegraphics[width=\textwidth]{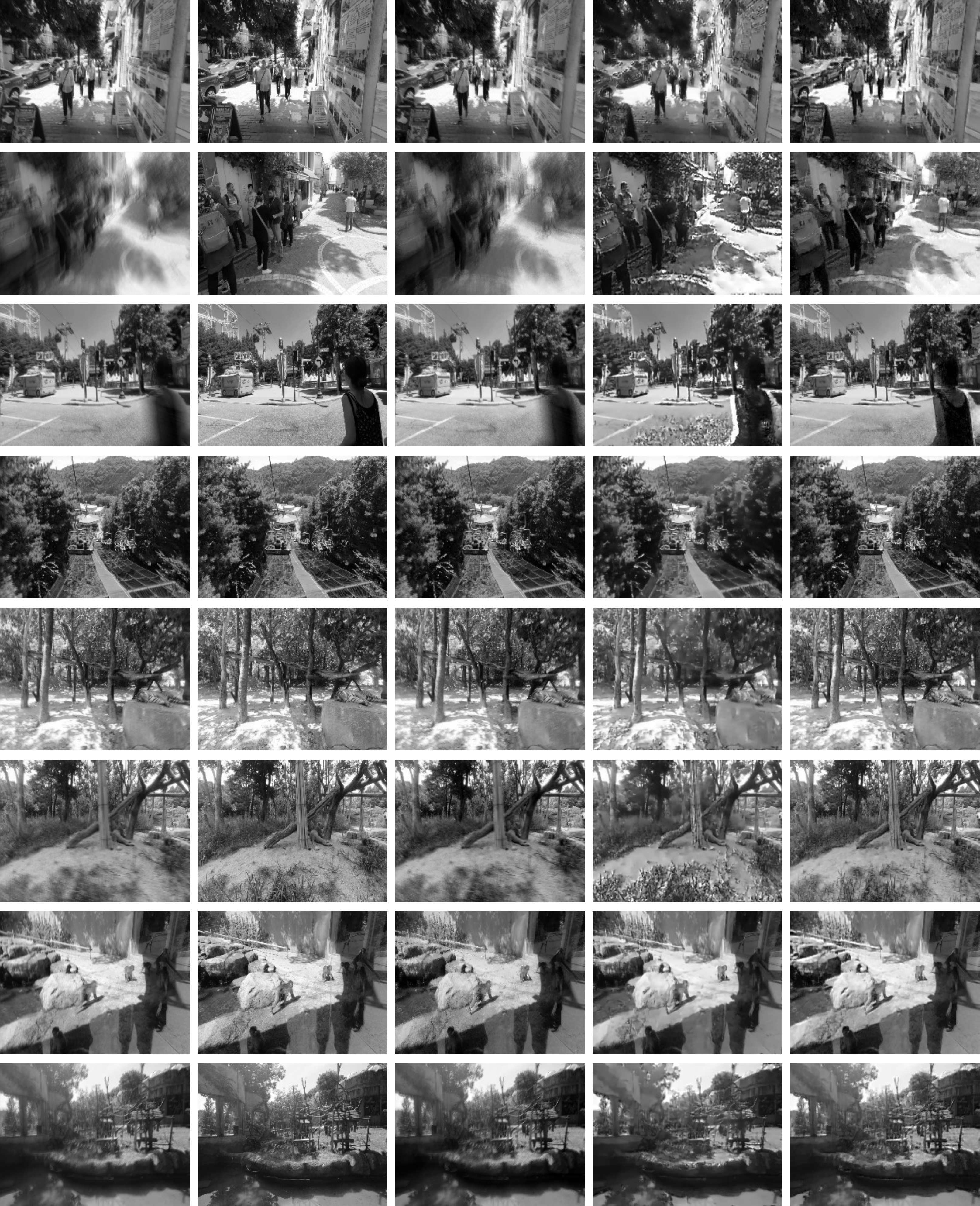}
    \begin{tabularx}{\textwidth}{Y Y Y Y Y}
        Input & Ground Truth & EDI~\cite{pan2019bringing} & eSL-Net~\cite{wang2020event} & Ours 
    \end{tabularx}
    \vspace{-0.2in}
    \caption{Qualitative visualization on the REDS~\cite{Nah_2019_CVPR_Workshops_REDS} dataset.}
    \label{fig:reds_3}
\end{figure*}

\begin{figure*}
    \centering
    \includegraphics[width=\textwidth]{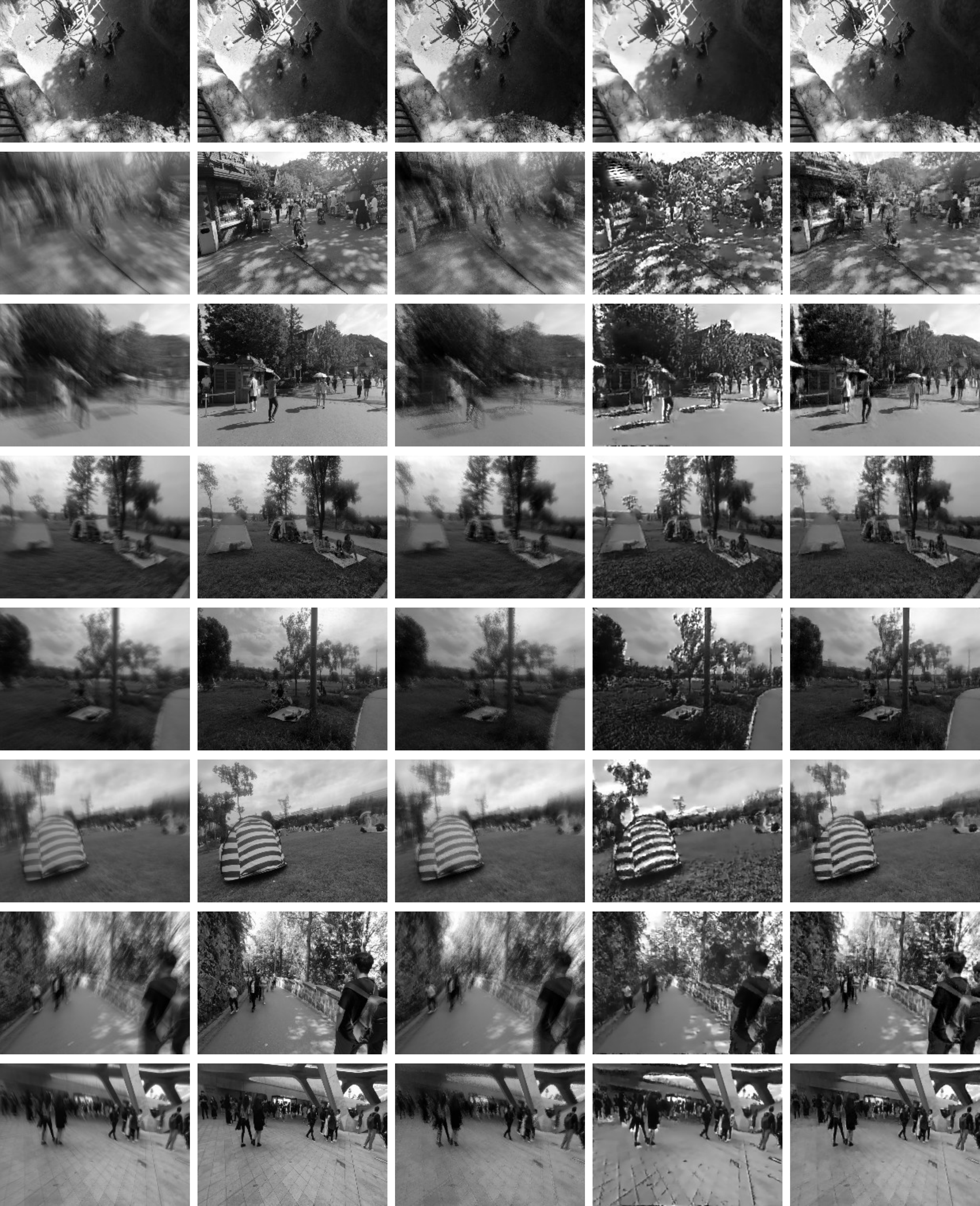}
    \begin{tabularx}{\textwidth}{Y Y Y Y Y}
        Input & Ground Truth & EDI~\cite{pan2019bringing} & eSL-Net~\cite{wang2020event} & Ours 
    \end{tabularx}
    \vspace{-0.2in}
    \caption{Qualitative visualization on the REDS~\cite{Nah_2019_CVPR_Workshops_REDS} dataset.}
    \label{fig:reds_4}
\end{figure*}

\begin{figure*}
    \centering
    \includegraphics[width=\textwidth]{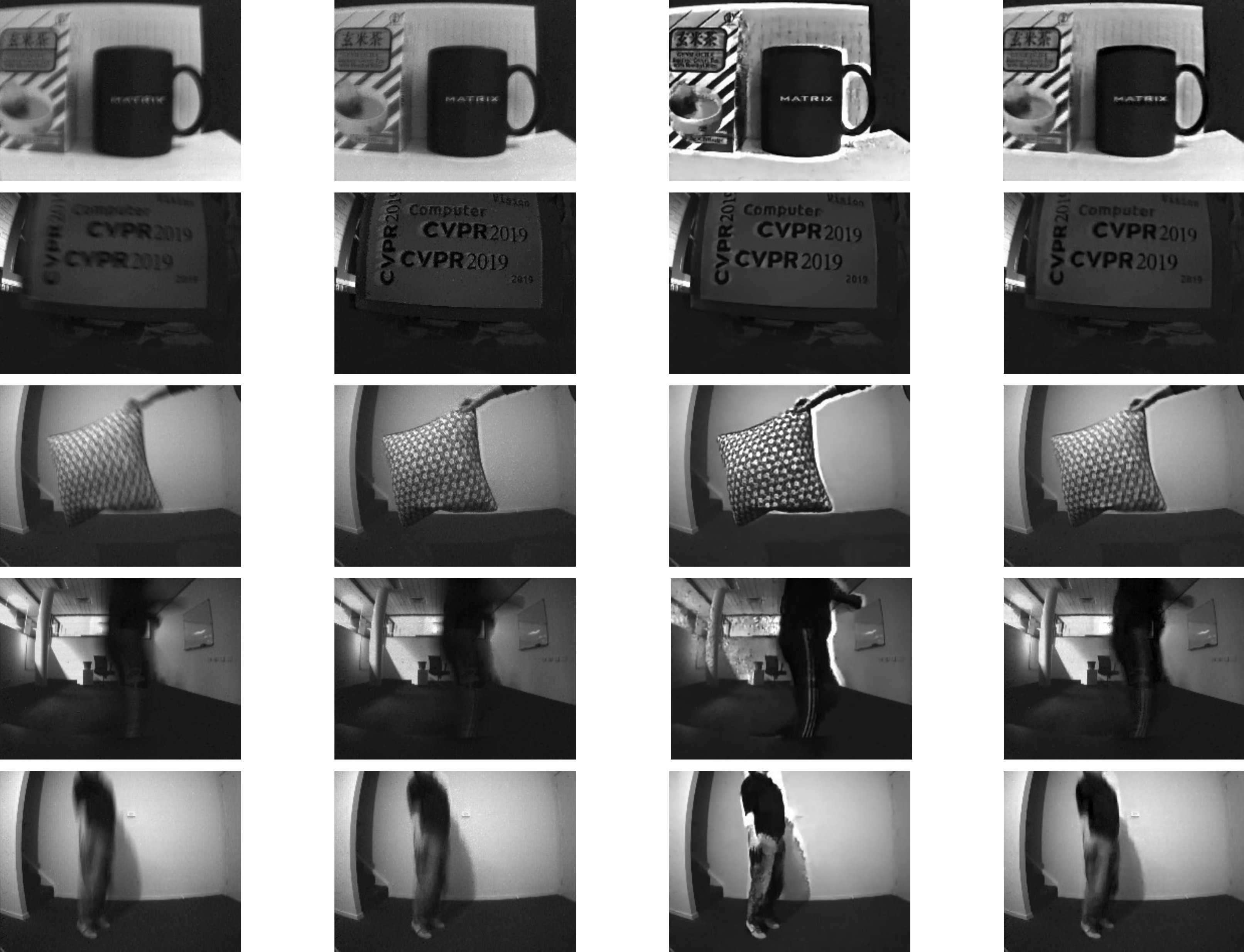}
    \begin{tabularx}{\textwidth}{Y Y Y Y}
        Input~~~~~~~~~~~ & EDI~\cite{pan2019bringing} & ~~~~~~eSL-Net~\cite{wang2020event} & ~~~~~~~Ours 
    \end{tabularx}
    \vspace{-0.2in}
    \caption{Qualitative visualization on real captures by Pan et al.~\cite{pan2019bringing}.}
    \label{fig:real_2}
\end{figure*}

\clearpage

{\small
\bibliographystyle{ieee_fullname}
\bibliography{egbib}
}
\end{document}

%% file: 00-Abstract.tex
\begin{abstract}
    A camera begins to sense light the moment we press the shutter button. During the exposure interval, relative motion between the scene and the camera causes motion blur, a common undesirable visual artifact. This paper presents \OurModel, which converts a blurry image into a sharp video represented as a parametric function from time to intensity. \OurModel~leverages events as an auxiliary input. We discuss how to exploit the temporal event structure to construct the parametric bases. We demonstrate how to train a deep learning model to predict the function coefficients. To improve the appearance consistency, we further introduce a refinement module to propagate visual features among consecutive frames. Compared to state-of-the-art event-enhanced deblurring approaches, \OurModel~generates smoother and more realistic results. The implementation of \OurModel~is available at \href{https://github.com/chensong1995/E-CIR}{https://github.com/chensong1995/E-CIR}.
\end{abstract}
\vspace{-0.2in}

%% file: 01-Introduction.tex
\section{Introduction}
\label{Sec:Introduction}
The shutter speed, or the length of the exposure interval, controls how much light reaches the image sensor from the environment. If the exposure interval is too short, the camera only has the time to capture very few photons. Consequently, the resulting image is not only unilluminated but also lacks fine details. On the other hand, if the exposure interval is too long, the relative motion between the scene and the camera may potentially be very significant. The resulting image is then the temporal average of a moving trajectory, causing blurry artifacts. 
Traditionally, it is presumed that any motion during the exposure interval, including both camera shake and subject movement, is unwanted and should therefore be removed. Over the past several decades, researchers have studied extensively how to convert a blurry image into a sharp one~\cite{richardson1972bayesian, fish1995blind, krishnan2009fast, joshi2009image, levin2007image, 713236, shan2008high, fergus2006removing, xu2013unnatural, xu2010two, 6909768, babacan2012bayesian, kupyn2018deblurgan, kupyn2019deblurgan}. It is only until recently when several works that reconstruct the complete motion trajectory have received profound attention~\cite{jin2018learning, purohit2019bringing}. These works introduce algorithms that convert a blurry image into a sharp video describing the exact movement that causes the blurry artifact.

\begin{figure}
    \centering
    \includegraphics[width=0.95\columnwidth]{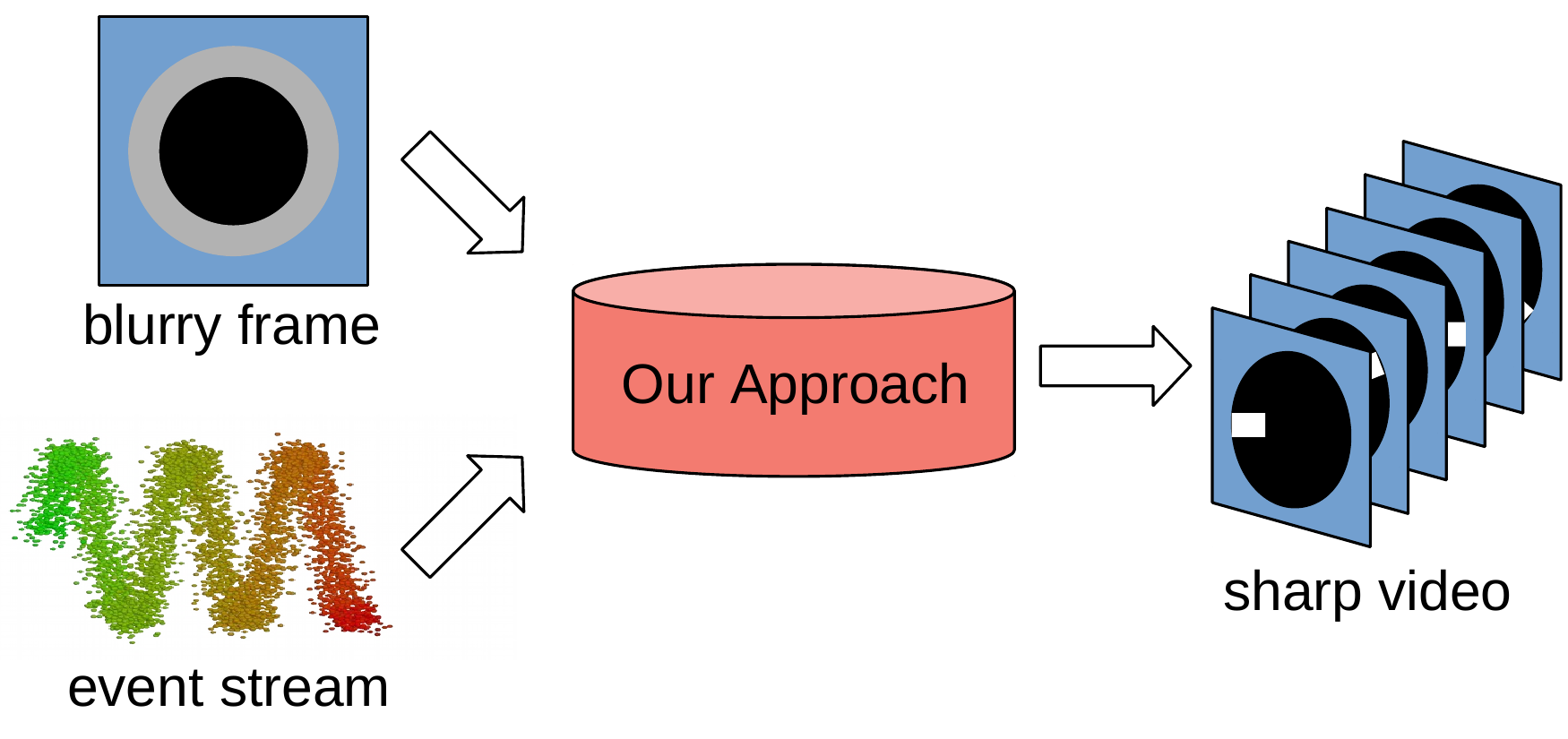}
    \caption{Problem Description. In this imaginary scene, we place a white square along the edge of a black disk. The image taken by the conventional camera is blurry because the disk rotates at a fast speed. It is as if the perimeter of the disk somehow grows into a gray collar. During the exposure interval, the event sensor produces a spiral of events. Our approach takes the blurry frame and the events as input and produces a sharp video sequence as output. The output video explains the motion blur by entailing the complete motion trajectory of the rotating disk.}
    \label{fig:problem}
    \vspace{-0.2in}
\end{figure}

Sharp video reconstruction is an ill-posed problem because there are infinitely many motion trajectories whose temporal averages correspond to the same blurry frame. To compensate for the ambiguity, previous works~\cite{pan2019bringing, pan2020high, jiang2020learning, lin2020learning, wang2020event, zhang2021fine, shang2021bringing, han2021evintsr, xu2021motion} exploit event data as an auxiliary input, which provides additional information during the exposure interval at a finer temporal resolution, as shown in Figure~\ref{fig:problem}. 
Even with the event input, difficult challenges remain. The events fail to capture the complete motion information. The video reconstruction quality is determined not only by the appearance of each individual frame but also the temporal smoothness. The immense density of events creates another obstacle for effective and efficient processing. The success of video deblurring depends on how the blurry image, the events, and priors about video sequences are integrated together. This calls for suitable video representations and prediction algorithms.

This paper makes fundamental contributions in video representations and methodologies for recovering accurate and temporally consistent videos. Specifically, we propose a continuous video representation whose coefficients are highly interpretable and easy to learn, due to their strong correlation to the events. For every pixel $(x, y)$, we represent its intensity as a parametric polynomial function $\mathbf{L}_{xy}(t)$, allowing us to render the sharp image at any given timestamp $t$ during the exposure interval. We show how to choose the polynomial bases such that the derivative of $\mathbf{L}_{xy}(t)$ interpolates the significant intensity changes. We also demonstrate how to train a deep neural network that regresses the polynomial coefficients. Instead of processing the video as a volume and implicitly encoding motions in convolutional filters, our approach explicitly asks the model to elaborate the motions that have already been described by the events. To further polish the frame quality, we introduce a refinement module that propagates the visual features among consecutive frames, which can be trained in an end-to-end manner with the rest of the model. The proposed regress-and-refine paradigm nicely combines the strength of recurrent modules for enforcing temporal smoothness and the strength of regression for drifting avoidance. 

We quantitatively evaluate our method on the synthetic REDS dataset~\cite{Nah_2019_CVPR_Workshops_REDS}. In terms of reconstruction quality, \OurModel~achieves an MSE of 0.114, representing a 37.4\% improvement from state-of-the-art algorithms. We also present a qualitative evaluation on the real captures provided by Pan et al.~\cite{pan2019bringing}. Compared with baseline approaches, our method is less noisy, more realistic, and temporally smoother. 

In summary, our key contributions are:

\begin{enumerate}
    \item We represent a video by per-pixel parametric polynomials. We discuss why this representation integrates easily with the event mechanism by showing the parallelism between function derivatives and events.
    \item From a blurry image and its associated events in the exposure interval, we demonstrate how to use a deep learning model to predict a sharp video represented by the proposed parametric polynomials.
    \item To overcome the limitations of the polynomial representation, we discuss how to formulate a refinement objective and encourage the temporal propagation of sharp visual features.
    \item We provide source code and documentation for converting the original REDS dataset into the event format. This clears the vagueness of the evaluation dataset in previous works and establishes an open-source benchmark for future comparisons.
\end{enumerate}

%% file: 02-RelatedWork.tex
\section{Related Work}
\label{Sec:Related-Work}

\subsection{Event-Enhanced Deblurring}
First commercialized in 2006~\cite{1696265}, event cameras are an emerging type of vision sensor that models the environment evolution as intensity changes and represents the scene as \textit{events}. Each event is a 4-tuple $(x, y, t, p)$ that contains the location, time, and polarity of an intensity change. This simple representation allows event cameras to support a fast data rate (up to 1 MHz), orders of magnitude higher than the frame rate of conventional cameras. The density of events during the exposure interval provides valuable motion information to explain the blurred image.

Pan et al. propose the Event-based Double Integral (EDI) model~\cite{pan2019bringing, pan2020high} that analytically reconstructs a high frame-rate sharp video from a blurry frame and its associated events. Jiang et al.~\cite{jiang2020learning} formulate a Maximum-a-Posteriori problem and solve for the latent sharp images under the Markov assumption with the help of deep neural networks. Lin et al.~\cite{lin2020learning} 
believe it is inadequate to calculate the intensity residual between sharp and blurry frames directly from the event threshold and propose to predict the intensity residual using deep learning. Meanwhile, the structural similarity between the EDI model and the blur kernel formulation has inspired Wang et al.~\cite{wang2020event} to represent sharp images as sparse codes in a learnable dictionary and optimize them using an iterative network. Shang et al.~\cite{shang2021bringing} assume that the input sequence contains a mixture of blurry and sharp frames and propose to wrap sharp frames to deblur the blurry frame. Zhang et al.~\cite{zhang2021fine} emphasize the temporal correlation among consecutive frames and design a multi-patch convolutional LSTM to exploit such correlation. Han et al.~\cite{han2021evintsr} extend this idea by modeling the intensity residual between neighboring sharp frames. Xu et al.~\cite{xu2021motion} also identify the importance of temporal correlation and propose to utilize the optical flow estimation instead.

Closely related to deblurring, event-enhanced frame interpolation has also attracted increasing attention~\cite{tulyakov2021time, Paikin_2021_CVPR}. While both tasks aim at constructing a high frame-rate video, frame interpolation methods typically assume the input frames are free of motion blur. Several works have attempted to reconstruct a high frame-rate video directly from the events without the conventional frame input~\cite{Rebecq19cvpr, Rebecq19pami, 9337171} as well. However, these event-only methods are less robust than their dual-input counterparts~\cite{zhang2021fine, lin2020learning}.

\subsection{Video Representation}
To the best of our knowledge, most existing works in computer vision process videos as discrete collections of frames. The only exception is Vid-ODE~\cite{park2020vid}, which represent videos by continuous latent states. The latent state can be evaluated at any given timestamp, allowing the video to be rendered with an infinitely high frame rate.

With the help of per-pixel parametric polynomials, our proposed representation also supports infinitely high rate rendering and enjoys two additional advantages. First, the polynomial bases are chosen to closely mimic the event mechanism, which makes the algorithm robust to domain differences between synthetic training data and real testing data. Second, the polynomial coefficients are more interpretable than the latent code hidden inside a deep network. This allows humans to easily explain and debug the model.

%% file: 03-Preliminary.tex
\section{Preliminaries}
\label{Sec:Preliminary}
\subsection{Event Camera Model}
\label{Sec:Preliminary:EventCameraModel}
Let $\mathbf{L}_{xy}(t)$ be the latent intensity of pixel $(x, y)$ at time $t$. In the natural logarithmic space, the temporal contrast between $t_{\text{ref}}$ and $t$ is given by~\cite{1696265}:
\begin{equation}
    \Delta ln[\mathbf{L}_{xy}(t)] = ln[\mathbf{L}_{xy}(t)] - ln[\mathbf{L}_{xy}(t_\text{ref})]
\end{equation}
where $t_{\text{ref}}$ denotes the timestamp of the last event associated with pixel $(x, y)$. The magnitude of $\Delta ln[\mathbf{L}_{xy}(t)]$ determines whether the hardware produces an event. Let $(x, y, t, p)$ denote an event, where $p \in \{-1, +1\}$ is the polarity of the intensity change~\cite{1696265}:
\begin{equation}
    p = \begin{cases} 
            +1 & \Delta ln[\mathbf{L}_{xy}(t)] \geq c^+ \\
            0 \text{ (no event)} & c^- < \Delta ln[\mathbf{L}_{xy}(t)] < c^+ \\
            -1 & \Delta ln[\mathbf{L}_{xy}(t)] \leq c^-
        \end{cases}
    \label{eq:polarity-definition}
\end{equation}
Here, $c^+$ and $c^-$ are thresholds controlling the sensitivity of positive and negative events, respectively. It is commonly assumed that $c^+$ and $c^-$ are stochastic variables~\cite{4444573, Rebecq18corl}. 

During an exposure interval $[-\frac{T}{2}, \frac{T}{2}]$ with length $T$, let $\mathbf{B}_{h\times w} = \{\mathbf{B}_{xy}\}$ be the blurry output from the conventional camera. The relation between the blurry frame and the latent frames is given by temporal averaging~\cite{Rebecq18corl, pan2019bringing}:
\begin{equation}
    \mathbf{B}_{xy} = \frac{1}{T} \int_{-\frac{T}{2}}^{\frac{T}{2}} \mathbf{L}_{xy}(t) dt
    \label{eq:blurry-definition}
\end{equation}

\subsection{Task Description}
\label{Sec:Preliminary:TaskDescription}
The input to the task has two components:
\begin{enumerate}
    \item The blurry intensity $\mathbf{B}_{xy}$ for all pixels $(x, y)$;
    \item A collection of events during the exposure interval $\{e_i = (x_i, y_i, t_i, p_i) | -\frac{T}{2} \leq t_i \leq \frac{T}{2}\}$.
\end{enumerate}

Given an arbitrary timestamp $t \in [-\frac{T}{2}, \frac{T}{2}]$, the goal of the task is to construct the corresponding latent frame $\mathbf{L}_{h \times w}(t) = \{\mathbf{L}_{xy}(t)\}$.

\subsection{Challenges}
\label{Sec:Preliminary:Challenges}
Intensity reconstruction is a highly ill-posed problem because there are infinitely many motion trajectories whose temporal averages correspond to the same blurry frame. Specifically, we face three challenges:

First, the events fails to capture complete motion information during the exposure interval. 
Equation~(\ref{eq:polarity-definition}) states that when an event happens, the magnitude of $\Delta ln[\mathbf{L}_{xy}(t)]$ is greater than the event threshold. It remains unclear exactly how much $\Delta ln[\mathbf{L}_{xy}(t)]$ exceeds the threshold. Imagine a scene with two edges moving in the same pattern. The first edge has a strong contrast to the background and generates significant intensity changes for its movement. The second edge has a weak contrast to the background and generates small intensity changes. Suppose these two sets of intensity changes both exceed the event threshold. This means the number of events generated by the camera is determined only by the edge length. These two edges will yield the same number of events, as long as their lengths are equal, even though the absolute change in intensity of the first edge is several times higher than that of the second edge. 

Second, the reconstruction quality is determined not only by the appearance of each individual frame but also the temporal smoothness. A naive model that independently optimizes each latent frame's quality may lead to unrealistic motion trajectories and cause frequent jitter. We refer readers to the supplementary animations for a demonstration of how some of our baseline approaches fail to solve this issue.

Third, the event format is incompatible with popular deep learning models. One possible remedy is to aggregate the events into a histogram.
This method ignores the event camera model and trains a network as if the inputs are merely some unexplained features. It remains unclear how to properly instill human knowledge about the bio-inspired event mechanism, such as the correlation between an event and an intensity change, into the model design.

While previous works fail to address some or all of these challenges, the next section discusses how the proposed \OurModel~handles them effectively.

%% file: 04-Method.tex
\section{Method}
\label{Sec:Method}

\subsection{Parametric Intensity Function}
\label{Sec:Method:ParametricIntensityRepresentation}
\begin{figure}
    \centering
    \includegraphics[width=0.49\columnwidth]{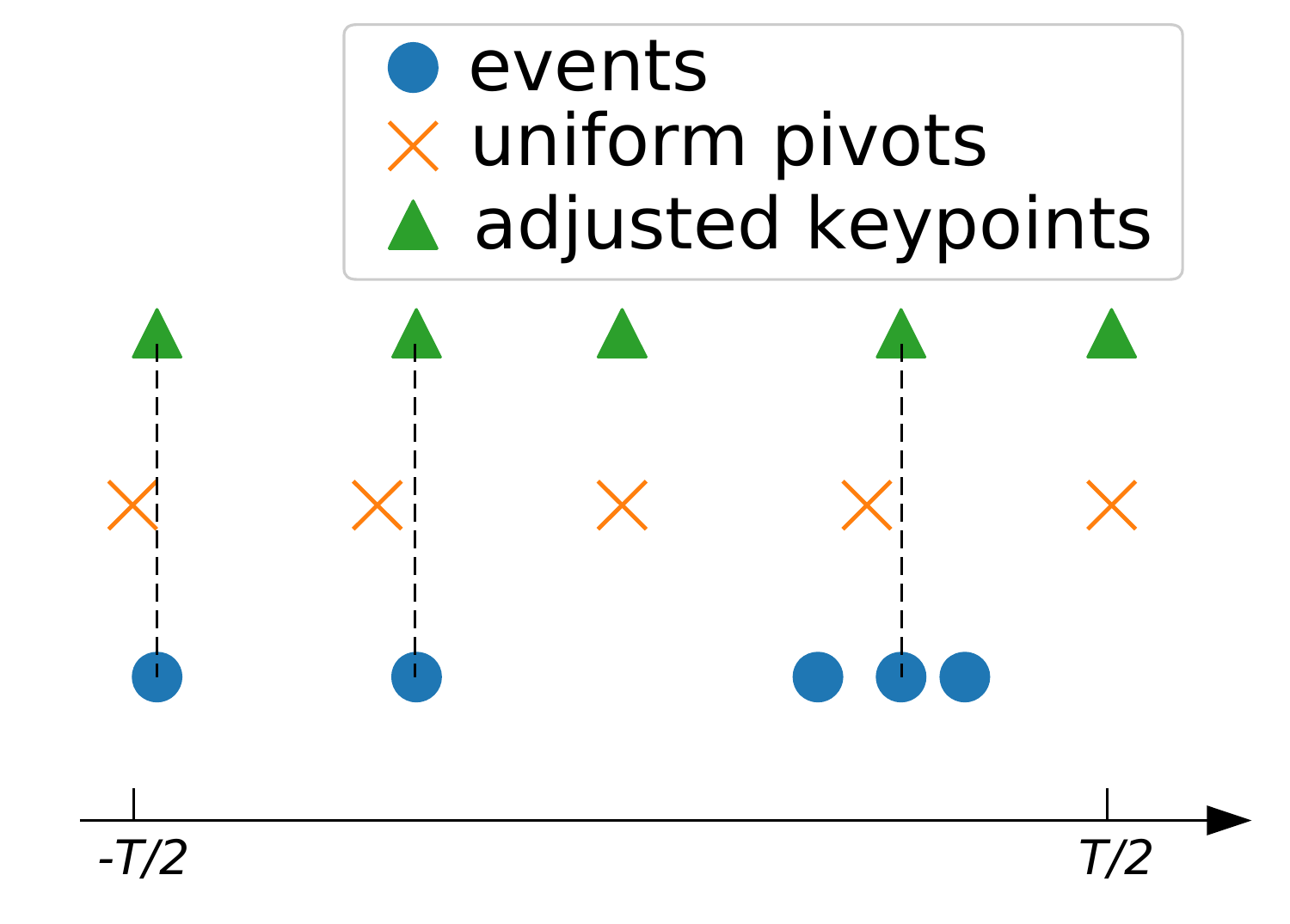} \includegraphics[width=0.49\columnwidth]{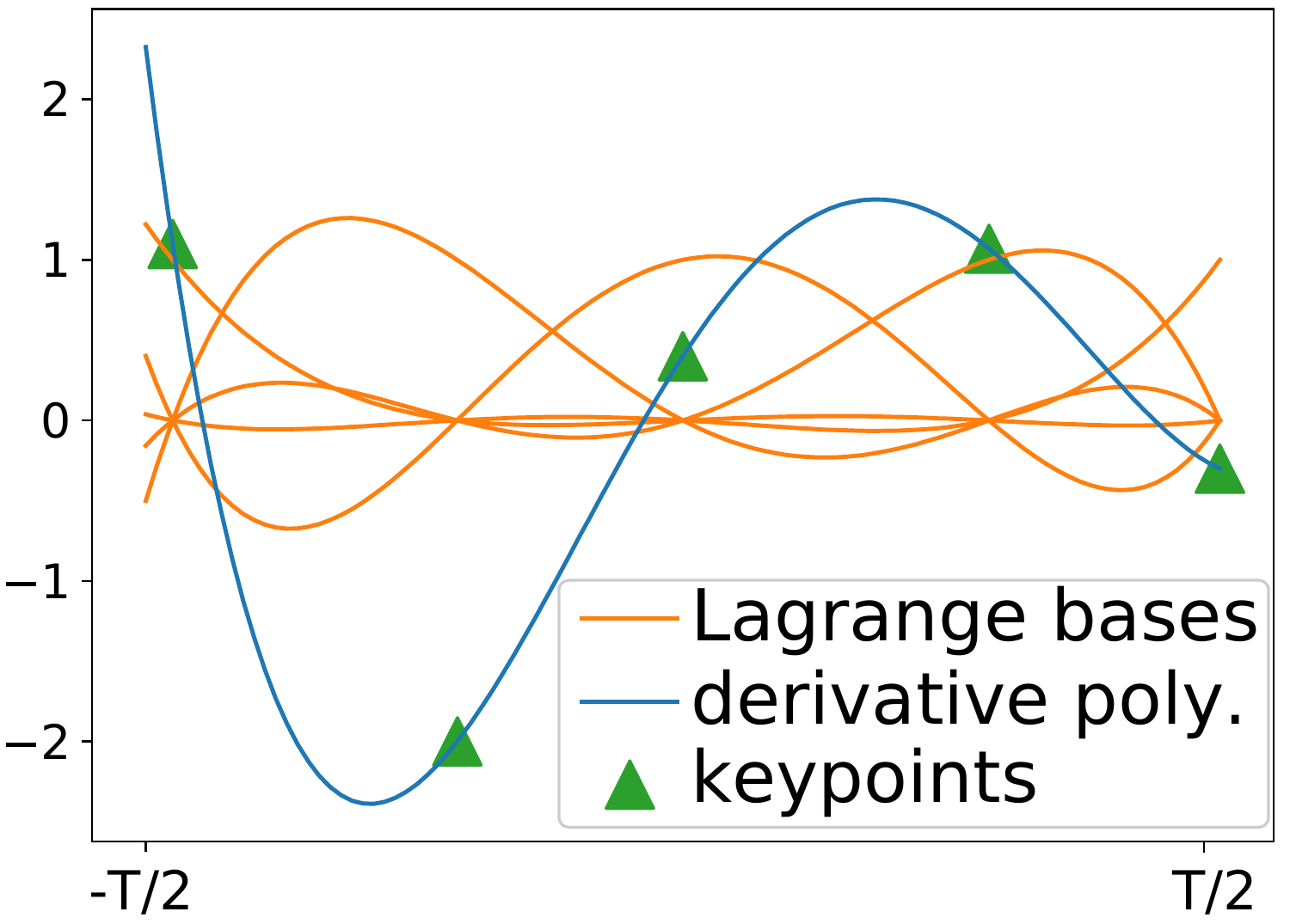}
    
    \begin{tabularx}{\columnwidth}{Y Y}
        (a) & (b)
    \end{tabularx}
    \caption{Keypoint Selection Algorithm. (a) Imagine there are $n=5$ different events scattered across the exposure interval represented by blue circles. We first sample $n$ evenly spaced pivots from the exposure interval $[-\frac{T}{2}, \frac{T}{2}]$ (the orange crosses). We then shift the pivots to their closest events and obtain $n$ keypoints (green triangles). This selection scheme not only caters to the temporal event structure but also provides support to regions not covered by events.  (b) At each keypoint, exactly one of the Lagrange bases has a value of 1, while all other Lagrange bases have a value of 0. We use a neural network to predict the value of $\frac{d\mathbf{L}_{xy}}{dt}(t)$ at these $n$ keypoints. Their interpolation gives a degree-$(n-1)$ polynomial, which is used to recover the primitive intensity signal $\mathbf{L}_{xy}(t)$ through indefinite integral. Under the Lagrangian representation, the polynomial coefficients coincide with predicted derivatives.}
    \label{fig:keypoint-selection}
    \vspace{-0.2in}
\end{figure}
\begin{figure*}
    \centering
    \includegraphics[width=0.95\textwidth]{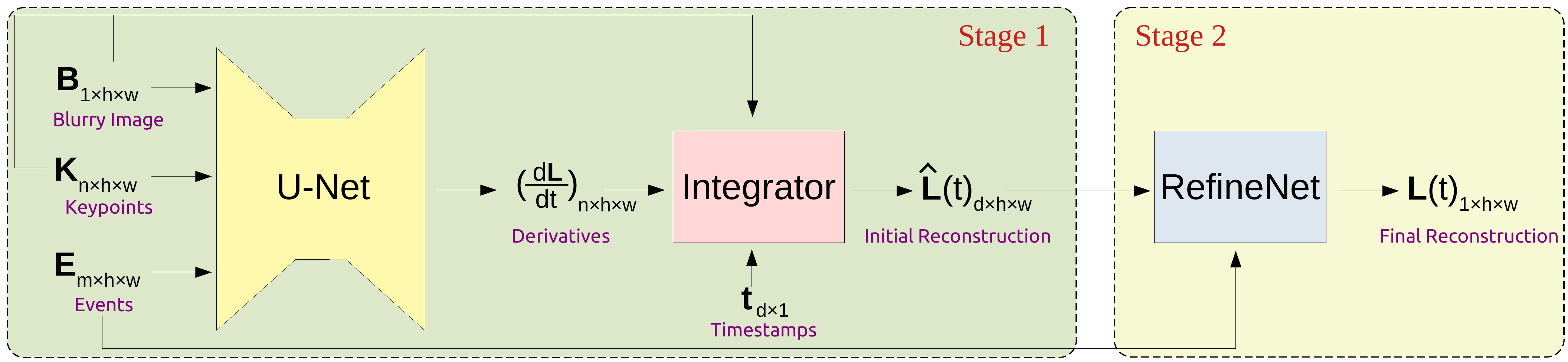}
    \caption{The overall pipeline. We use the U-Net~\cite{ronneberger2015u} model to regress the polynomial coefficients. The network takes three inputs: the blurry frame $\mathbf{B}$, the keypoint timestamps $\mathbf{K}$, and the event histogram $\mathbf{E}$. The network then outputs the intensity derivatives $\frac{d\mathbf{L}}{dt}$. Given an arbitrary timestamp $t$, the integrator follows Equation~(\ref{eq:primitive-recover}) and calculates the initial frame reconstruction $\mathbf{\hat{L}}(t)$. The refinement module takes $\mathbf{\hat{L}}(t)$ and $\mathbf{E}$ as input and outputs the final frame reconstruction $\mathbf{L}(t)$. In this figure, $m, n, d, h, w$ represent the number of histogram bins, the number of keypoints, the number of frames in the output video, the frame height, and the frame width, respectively.}
    \label{fig:model}
    \vspace{-0.2in}
\end{figure*}
For each pixel $(x, y)$, we propose to approximate the function $\mathbf{L}_{xy}(t)$ as a continuously differentiable mapping from the time domain to the intensity domain: $[-\frac{T}{2}, \frac{T}{2}] \rightarrow [0, 1]$. Inspired by the Taylor's theorem, we parameterize the mapping as a degree-$n$ polynomial. Let $\alpha_0, \alpha_1, \alpha_2, \cdots, \alpha_n$ be the $n+1$ polynomial coefficients. The simplest parameterization uses the standard polynomial bases:
\begin{equation}
    \mathbf{L}_{xy}(t) = \alpha_0 + \alpha_1 t + \alpha_2 t^2 + \cdots + \alpha_n t^n
\end{equation}

Most real-life videos do not have frequent oscillations. This means a realistic intensity curve does not have very high-order derivative information. For a small change in $t$, the variation in $\mathbf{L}_{xy}(t)$ should not be too big. The fact that $|\alpha_0| \approx |\alpha_1| > |\alpha_2| > \cdots \gg |\alpha_n|$ suggests that the standard representation is prone to numerical issues since high-order coefficients are expected to be very close to zero.

The temporal derivative of $\mathbf{L}_{xy}(t)$ reveals how the intensity changes across time. The events associated with pixel $(x, y)$ provide a set of timestamps where the intensity change considerably. The derivatives at these timestamps are expected to have significant magnitudes. The key idea of our proposed parameterization is to interpolate the temporal derivative of the intensity signal at event timestamps. The number of events associated with each pixel is different, presenting a challenge to efficient computation. To address this issue, we extract a fixed number of $n$ keypoints for each pixel, regardless of how many events the pixel initially possesses. The details of our keypoint extraction algorithm are presented in Figure~\ref{fig:keypoint-selection}(a). This algorithm ensures the selected keypoints are in correspondence to the event timestamps and as distant to each other as possible. The use of the uniform pivots further establishes spatial consistency in the keypoint choices among different pixels. Let the set of $n$ keypoints for pixel $(x, y)$ be:
\begin{equation}
    \mathcal{K}_{xy} = \{ (t_i, \frac{d\mathbf{L}_{xy}}{dt}(t_i)) | 1 \leq i \leq n\}
\end{equation}
where $-\frac{T}{2} \leq t_1 < t_2 < \dots < t_n \leq \frac{T}{2}$. As shown in Figure~\ref{fig:keypoint-selection}(b), we parameterize the intensity derivative $\frac{d\mathbf{L}_{xy}}{dt}(t)$ as the polynomial interpolation of these $n$ keypoints:
\begin{equation}
    \frac{d\mathbf{L}_{xy}}{dt}(t) = \sum_{i=1}^{n} \frac{d\mathbf{L}_{xy}}{dt}(t_i) \cdot \mathbf{\beta}_{xyi}(t)
    \label{eq:derivative-polynomial}
\end{equation}
Here, $\mathbf{\beta}_{xyi}(t), 1 \leq i \leq n$ are the Lagrange bases of degree $n-1$:
\begin{align*}
    \mathbf{\beta}_{xyi}(t) &= \frac{(t - t_1)\cdots(t - t_{i-1})(t - t_{i+1})\cdots(t - t_n)}{(t_i - t_1)\cdots(t_i - t_{i-1})(t_i - t_{i+1})\cdots(t_i - t_n)} 
\end{align*}

The Lagrange bases have the characteristic that $\mathbf{\beta}_{xyi}(t_i) = 1$ and $\forall j \neq i, \mathbf{\beta}_{xyj}(t_i) = 0$. This ensures the continuous function $\frac{d\mathbf{L}_{xy}}{dt}(t)$ passes through the $n$ discrete keypoints $(t_i, \frac{d\mathbf{L}_{xy}}{dt}(t_i))$, where $1 \leq i \leq n$.

We can then recover the primitive intensity signal $\mathbf{L}_{xy}(t)$ from its derivative $\frac{d\mathbf{L}_{xy}}{dt}(t)$ by taking the indefinite integral:
\begin{equation}
     \mathbf{L}_{xy}(t) = \int \frac{d\mathbf{L}_{xy}}{dt}(t) dt + a_{xy}
     \label{eq:primitive-recover}
\end{equation}
where $a_{xy}$ is a constant that can be solved from Equation~(\ref{eq:blurry-definition}).

Compared to the conventional frame-based representation, the main advantage of the  proposed polynomial representation is that it closely mimics the event mechanism. We leverage event timestamps to construct the polynomial bases and allow the polynomial coefficients to be interpreted as the intensity changes that trigger the input events. The regression target of our model, the polynomial coefficients, is therefore highly correlated to the input events. While the input events characterize the locations of the edge features, the output polynomial coefficients reveal exactly how significant the edges are. Section~\ref{sec:Evaluation:Analysis} presents an empirical verification of the advantage of our representation.

\subsection{Prediction Pipeline}
\label{Sec:Method:PredictionPipeline}

We illustrate the overall prediction pipeline in Figure~\ref{fig:model}, which consists of the initialization stage and the refinement stage. The initialization stage regresses the polynomial coefficients, evaluates predicted parametric functions, and obtains coarse video reconstruction. The refinement stage further polishes the details in the initial reconstruction by learning and enforcing motion priors. This methodology nicely combines the strength of motion priors for recovering temporally smooth videos and the strength of regressing the volumetric output to avoid drifting.  

\noindent\textbf{Initialization: Polynomial Coefficient Regression.} We assemble the $n$ keypoint timestamps associated with each pixel into an $n \times h \times w$ tensor $\mathbf{K}$, where $h \times w$ is the spatial frame resolution. Following eSL-Net~\cite{wang2020event}, we voxelize the event stream by creating an $m \times h \times w$ histogram tensor $\mathbf{E}$, where $m=40$ is the number of temporal bins. We adopt the U-Net~\cite{ronneberger2015u} architecture as the backbone prediction network. As shown in Figure~\ref{fig:model}, the U-Net takes the blurry frame, the keypoints, and the events as input and regresses the polynomial coefficients for $\mathbf{L}_{xy}(t)$ represented under the Lagrange bases of its derivative. Let $\mathbf{t} = \{t_i\}$ be the $d$-dimensional vector collecting all timestamps of interest. With the help of Equation~(\ref{eq:primitive-recover}), the predicted coefficients allow us to reconstruct $d$ initial frames $\{\mathbf{\hat{L}}(t_i)\}$. At training time, $d=14$ is set to the number of available ground-truth latent frames. 
At inference time, $d$ can be an arbitrary positive integer.

\begin{algorithm}
\caption{Refinement}\label{alg:refinement}
\textbf{Input:} Initial reconstruction: $\{\mathbf{\hat{L}}(t_i) | 1 \leq i \leq d\}$ \\
\textbf{Input:} All events: $\{(x, y, t, p) | -\frac{T}{2} \leq t \leq \frac{T}{2}\}$ \\
\textbf{Output:} Final reconstruction: $\{\mathbf{L}(t_i) | 1 \leq i \leq d\}$
\begin{algorithmic}[1]
\Loop{ $I_{max}$ iterations}
    \For{$i = 1 \textbf{ to } d-1$} \Comment{Residual Prediction}
        \State $\mathbf{E} \leftarrow$ \textsc{Voxelize}(events from $t_i$ to $t_{i+1})$
        \If{i == 1}
            \State $\mathbf{R}_i \leftarrow g_{\theta_1}^{\mathbf{R}}(\mathbf{E}, \mathbf{\hat{L}}(t_i), \mathbf{\hat{L}}(t_{i+1}),$
            \State ~~~~~~~~~~~~~~~~~ $\nabla\mathbf{\hat{L}}(t_i), \nabla\mathbf{\hat{L}}(t_{i+1}))$
        \Else
            \State $\mathbf{R}_i \leftarrow g_{\theta_2}^{\mathbf{R}}(\mathbf{R}_{i-1}, \mathbf{E}, \mathbf{\hat{L}}(t_i), \mathbf{\hat{L}}(t_{i+1}),$
            \State ~~~~~~~~~~~~~~~~~ $\nabla\mathbf{\hat{L}}(t_i), \nabla\mathbf{\hat{L}}(t_{i+1}))$
        \EndIf
    \EndFor
    \For{$i = 1 \textbf{ to } d$} \Comment{Apply Updates}
        \State $\mathbf{A_i} \leftarrow g_{\phi}^{\mathbf{A}}(\mathbf{\hat{L}}(t_i))$
        \State $\mathbf{D_i} \leftarrow \frac{\partial f(\mathbf{R}_1, \cdots, \mathbf{R}_{d-1}, \mathbf{\hat{L}}(t_1), \cdots, \mathbf{\hat{L}}(t_d), \mathbf{L}(t_1), \cdots, \mathbf{L}(t_d))}{\partial \mathbf{L}(t_i)}$
        \State $\mathbf{\hat{L}}(t_i) \leftarrow \mathbf{\hat{L}}(t_i) - \mathbf{A_i} \odot \mathbf{D_i}$
    \EndFor
\EndLoop
\For{$i = 1 \textbf{ to } d$} \Comment{Final Polishing}
    \State $\mathbf{L}(t_i) \leftarrow g_{\gamma}^{\mathbf{L}}(\mathbf{\hat{L}}(t_i))$
\EndFor
\end{algorithmic}
\end{algorithm}

\noindent\textbf{Refinement: Temporal Feature Propagation.} When played as a video, the initialization results show an accurate motion reconstruction. Different parts of the scene move according to their respectively trajectories. Nonetheless, we observe that the initial reconstruction occasionally fails to recover temporally consistent features. This initial temporal inconsistency is expected and addressed by the refinement. 


There are two factors that contribute to the temporal inconsistency. First, the polynomial function is a continuous signal that smooths sharp features and makes them visually blurry. Second, visual features may only move during a small part of the exposure interval. When the feature is actively moving, the reconstruction is usually sharp because there are input events in the spatial neighborhood depicting the edge locations. When the feature is not moving, however, the reconstruction becomes blurry due to the lack of associated events and difficulty for volumetric filters used in regression to capture motion priors that are critical for deblurring. 

In the refinement stage, we solve the first issue by optimizing the frames independent of polynomial formulation. We solve the second issue by encouraging visual features to propagate between consecutive frames (i.e., enforcing motion priors), a popular technique used extensively in video synthesis via optical flows~\cite{xu2021motion, caballero2017real}, residuals~\cite{han2021evintsr, li2019fast}, or deformable convolutional kernels~\cite{tian2018tdan, tian2020tdan}. Details of our refinement process are presented as Algorithm~\ref{alg:refinement}. Specifically, we use a recurrent network to predict the residual $\mathbf{R}_i$ between adjacent frames $\mathbf{L}(t_i)$ and $\mathbf{L}(t_{i+1})$. The inputs to the network include the previous residual $\mathbf{R}_{i-1}$ between frames $\mathbf{L}(t_{i-1})$ and $\mathbf{L}(t_i)$, the events from $t_i$ to $t_{i+1}$, the initial reconstructions $\mathbf{\hat{L}}(t_i)$ and $\mathbf{\hat{L}}(t_{i+1})$, as well as their spatial gradients $\nabla\mathbf{\hat{L}}(t_i)$ and $\nabla\mathbf{\hat{L}}(t_{i+1})$. Algorithm~\ref{alg:refinement} refers this recurrent network as $g_{\theta_1}^{\mathbf{R}}$ (for the residual prediction between the first two frames) and $g_{\theta_2}^{\mathbf{R}}$ (for the rest of residuals).

The recurrent architecture allows the residual to be gradually updated according to the relevant events and intensity reconstruction. We augment the inputs to include $\nabla\mathbf{\hat{L}}(t_i)=(\frac{d\mathbf{\hat{L}}(t_i)}{dx}, \frac{d\mathbf{\hat{L}}(t_i)}{dy})$ and $\nabla\mathbf{\hat{L}}(t_{i+1})=(\frac{d\mathbf{\hat{L}}(t_{i+1})}{dx}, \frac{d\mathbf{\hat{L}}(t_{i+1})}{dy})$. This is because both the spatial gradients and the temporal residuals are highly related to the edge features.

Consider the objective function in Equation~(\ref{eq:refinement-objective}), where $\mathcal{L}(\cdot, \cdot)$ represents the distance between two matrices. 

\begin{equation}
    f = \sum_{i=1}^{d-1} \mathcal{L}(\mathbf{L}(t_i) + \mathbf{R}_i, \mathbf{L}(t_{i+1})) + \lambda \sum_{i=1}^d \mathcal{L}(\mathbf{L}(t_i), \mathbf{\hat{L}}(t_i))
    \label{eq:refinement-objective}
\end{equation}

The free variables are the refined frames $\mathbf{L}(t_i)$'s. The first objective term ensures the refinement output follows the residual flow. The second term discourages the refinement from deviating too far from the initialization. The trade-off parameter $\lambda$ balances these two terms. We expect both terms to have small residuals throughout the optimization process and the final result to be in local proximity to the initial frames. This leads us to choose the L2-distance as $\mathcal{L}(\cdot, \cdot)$ for its numerical stability. As shown in Algorithm~\ref{alg:refinement}, we apply gradient descent to update the refinement result for $I_{max}$ iterations. Different pixels may have different step sizes for the update, which are predicted by a convolutional network referred to as $g_\phi^{\mathbf{A}}$. After the gradient descent terminates, we use another convolutional network $g_\gamma^{\mathbf{L}}$ to perform final polishing on each individual frame.

\subsection{Training Objective}
\label{Sec:Method:TrainingObjective}
\noindent\textbf{Derivative Loss}. We use $\mathcal{L}_\text{d}$ to supervise the output polynomial coefficients directly in the derivative domain:
\begin{equation}
    \mathcal{L}_\text{d} = |(\frac{d\mathbf{L}}{dt})_\text{gt} - (\frac{d\mathbf{L}}{dt})_\text{pred}|_1
\end{equation}


\noindent\textbf{Primitive Loss}. We first recover the primitive intensity signal from Equation~(\ref{eq:primitive-recover}) and then use $\mathcal{L}_\text{p}$ to supervise the polynomial coefficients indirectly in the primitive domain:
\begin{equation}
    \mathcal{L}_\text{p} = |\mathbf{L}_\text{gt} - \mathbf{L}_\text{pred}|_1
\end{equation}



\noindent\textbf{Refinement Loss}. We use $\mathcal{L}_\text{ref}$ to supervise the refinement output:
\begin{equation}
    \mathcal{L}_\text{ref} = \sum_t |\mathbf{L}_\text{gt}(t) - \mathbf{L}_\text{pred}(t)|_1
\end{equation}

\noindent\textbf{Residual Loss}. We use $\mathcal{L}_\text{res}$ to supervise the residual prediction in the refinement stage. Note that $\mathcal{L}_\text{res}$ uses weighted L1-norm with $\rho=5$ because the intensity residual between consecutive frames is sparse.
\begin{equation}
    \mathcal{L}_\text{res} = \sum_i (exp(\rho \cdot |\mathbf{R}_{i_\text{gt}}|_1) \odot |\mathbf{R}_{i_\text{gt}} - \mathbf{R}_{i_\text{pred}}|_1)
\end{equation}

\noindent\textbf{Total Objective}. The final training objective is the weighted sum of the losses introduced above:
\begin{equation}
    \mathcal{L}_\text{total} = \lambda_\text{d} \mathcal{L}_\text{d} +
                               \lambda_\text{p} \mathcal{L}_\text{p} +
                               \lambda_\text{ref} \mathcal{L}_\text{ref} +
                               \lambda_\text{res} \mathcal{L}_\text{res}
    \label{eq:objective}
\end{equation}
where $\lambda_\text{d}, \lambda_\text{p}, \lambda_\text{ref}, \lambda_\text{res}$ are constant trade-off factors. The U-Net and the prediction networks in the refinement stage ($g_{\theta_1}^{\mathbf{R}}, g_{\theta_2}^{\mathbf{R}}, g_{\phi}^{\mathbf{A}}, g_{\gamma}^{\mathbf{L}}$) are trained in an end-to-end manner.


%% file: 05-Evaluation.tex
\section{Evaluation}
\label{Sec:Evaluation}

\subsection{Datasets}
\label{Sec:Evaluation:Dataset}
REDS~\cite{Nah_2019_CVPR_Workshops_REDS} is a standard deblurring benchmark dataset designed for conventional cameras. The dataset contains 240 training videos and 30 validation videos and is publicly available under the CC BY 4.0 license. These videos are captured at 120 fps and are sharp and clear. We use the frame interpolation algorithm~\cite{niklaus2017video} to further increase the frame rate to 960 fps. After that, we convert the videos to grayscale and resize the frames to 240$\times$180, consistent with the DAVIS240\footnote{DAVIS240 is a popular camera that records grayscale conventional frames and events simultaneously.}~\cite{6889103} sensor resolution. We apply the ESIM~\cite{Rebecq18corl} simulator to generate events and synthesize blurry frames according to Equation~(\ref{eq:blurry-definition}) with an exposure interval of 120 milliseconds. We use least-squares to fit polynomial coefficients to the resized 960 fps sharp frames during each exposure interval. These coefficients are used to supervise network training. This process is an effort to reproduce the synthetic dataset described by Wang et al.~\cite{wang2020event}, who have released the model weights after training but have not disclosed data processing or training scripts. 

\subsection{Baseline Approaches}
\label{sec:Evaluation:Baseline}
\begin{table}[t]
    \centering
    \small
    \begin{tabular}{r|c c c}
        \hline
        Methods & MSE $\downarrow$ & PSNR $\uparrow$ & SSIM $\uparrow$ \\
        \hline
        EDI~\cite{pan2019bringing}           & 0.182 & 21.663 & 0.664 \\
        eSL-Net~\cite{wang2020event} (official) & 0.203 & 20.640 & 0.601 \\
        eSL-Net~\cite{wang2020event} (re-trained) & 0.201 & 20.748 & 0.646 \\
        \hline
        Ours                                 & \textbf{0.114} & \textbf{25.531} & \textbf{0.819}\\
        \hline
    \end{tabular}
    \caption{Quantitative evaluation on the REDS~\cite{Nah_2019_CVPR_Workshops_REDS} dataset.}
    \label{tab:reds_baseline}
    \vspace{-0.2in}
\end{table}
For quantitative evaluation, we compare our model with eSL-Net~\cite{wang2020event} using both the official weights released by Wang et al. and the weights re-trained on our synthetic data. We take EDI~\cite{pan2019bringing} as an additional baseline model. 
At the submission time, they are the only two approaches with open-source implementation available online.

For qualitative evaluation, we also visualize the results on real event captures provided by Pan et al.~\cite{pan2019bringing}. The blurriness of real captures comes from the physical sensor instead of simulated temporal averaging. Therefore, there is a lack of ``ground-truth'' sharp images, and we are unable to numerically evaluate the performance of each algorithm. 

\begin{figure*}
    \centering
    \includegraphics[width=\textwidth]{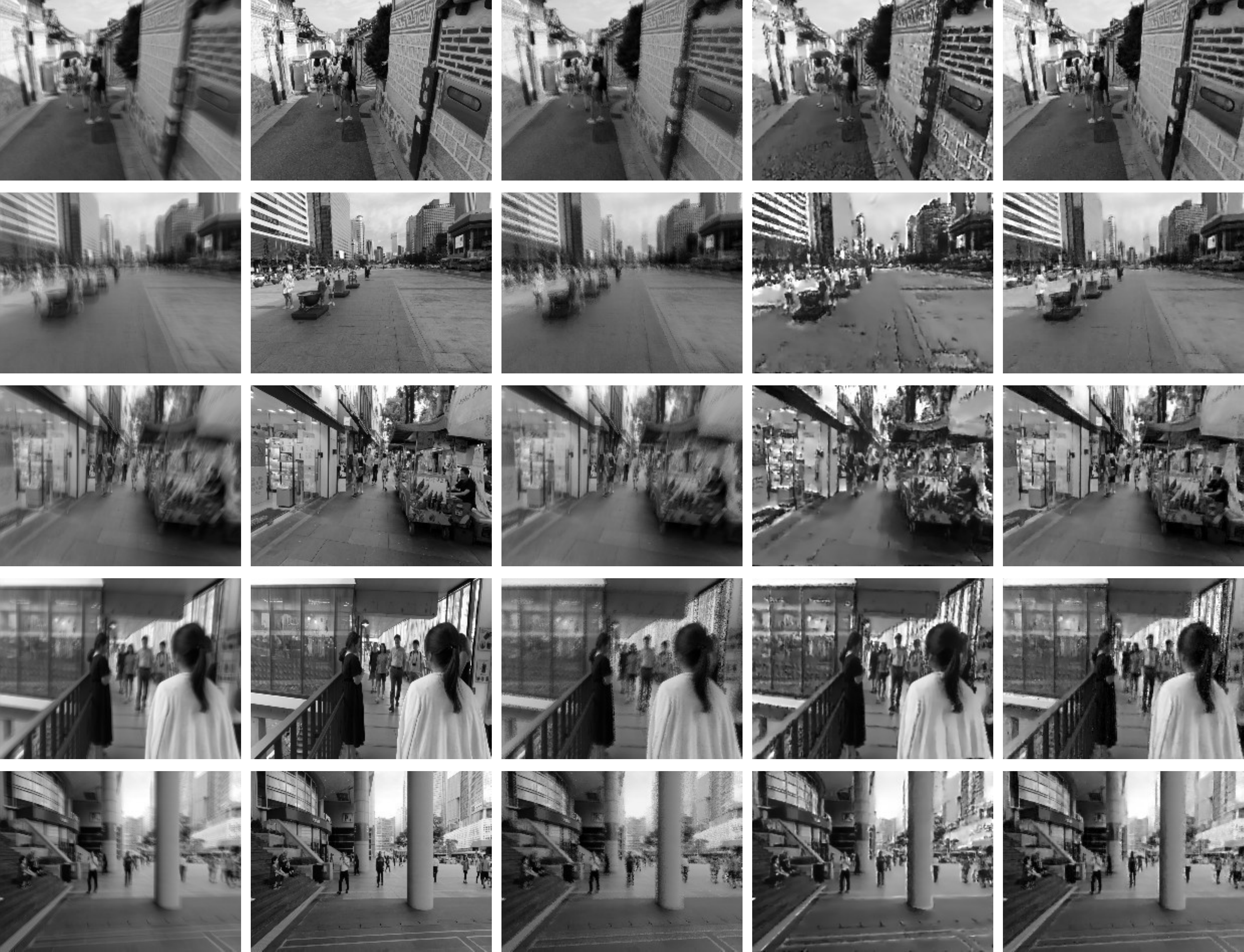}
    \begin{tabularx}{\textwidth}{Y Y Y Y Y}
        Input & Ground Truth & EDI~\cite{pan2019bringing} & eSL-Net~\cite{wang2020event} & Ours 
    \end{tabularx}
    \caption{Qualitative visualization on the REDS~\cite{Nah_2019_CVPR_Workshops_REDS} dataset.}
    \label{fig:reds_1}
\end{figure*}
\begin{figure*}
    \centering
    \includegraphics[width=\textwidth]{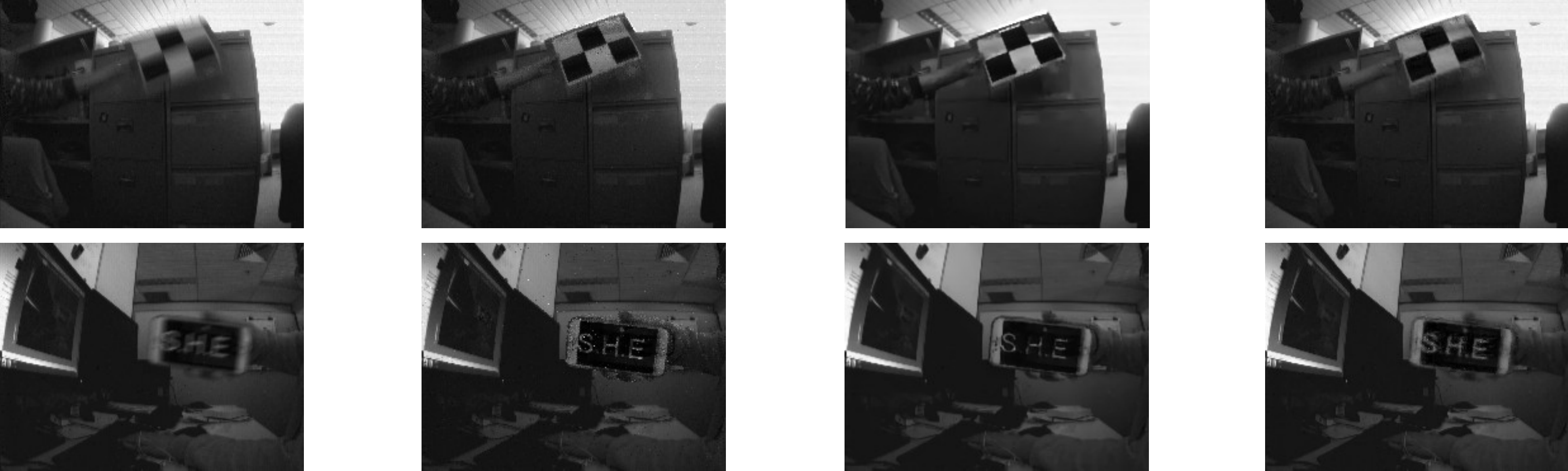}
    \begin{tabularx}{\textwidth}{Y Y Y Y}
        Input~~~~~~~~~~~ & EDI~\cite{pan2019bringing} & ~~~~~~eSL-Net~\cite{wang2020event} & ~~~~~~~Ours 
    \end{tabularx}
    \caption{Qualitative visualization on real captures by Pan et al.~\cite{pan2019bringing}.}
    \label{fig:real_1}
\end{figure*}
\begin{table*}[t]
    \centering
    \begin{tabular}{c | c c| c c | c c | c c c}
        \hline
        \multirow{2}{*}{Row} & \multicolumn{2}{c|}{Input Sources} & \multicolumn{2}{c|}{Video Format} & \multicolumn{2}{c|}{Stages} & \multicolumn{3}{c}{Performance Metrics} \\
        \cline{2-10}
        & Frame & Events & Poly. & Frame & Initialization & Refinement & MSE $\downarrow$ & PSNR $\uparrow$ & SSIM $\uparrow$ \\
        \hline
        1 & \redcross & \greencheck & \greencheck & \redcross & \greencheck & \redcross & 0.134 & 24.210 & 0.767 \\
        2 & \greencheck & \redcross & \greencheck & \redcross & \greencheck & \redcross & 0.180 & 21.721 & 0.654  \\
        3 & \greencheck & \greencheck & \greencheck & \redcross & \greencheck & \redcross & 0.125 & 24.807 & 0.787 \\
        4 & \greencheck & \greencheck & \redcross & \greencheck & \greencheck & \redcross & 0.136 & 23.504 & 0.723 \\
        \hline
        5 & \greencheck & \greencheck & \greencheck & \redcross & \greencheck & \greencheck  & \textbf{0.114} & \textbf{25.531} & \textbf{0.819} \\
        \hline
    \end{tabular}
    \caption{On the REDS~\cite{Nah_2019_CVPR_Workshops_REDS} dataset, we use ablation studies to demonstrate the importance of using a dual-stream input, the power of the polynomial representation, and the strength of the refinement module.}
    \label{tab:reds_ablation}
    \vspace{-0.2in}
\end{table*}

\subsection{Training Details}
\label{sec:Evaluation:Training}
The trade-off parameters for the derivative ($\mathcal{L}_\text{d}$), primitive ($\mathcal{L}_\text{p}$), refinement ($\mathcal{L}_\text{ref}$), and residual ($\mathcal{L}_\text{res}$) losses are 1, 10, 10, and 0.5, respectively. We use the Adam~\cite{kingma2014adam} optimizer with a batch size of 96 and train the network for 50 epochs. The initial learning rate is 0.0001 and is halved at the end of the 20$^{\text{th}}$ and the 40$^{\text{th}}$ epochs. The degree of the polynomial functions we use is $n=10$.

\subsection{Analysis of Results}
\label{sec:Evaluation:Analysis}

\noindent\textbf{Baseline Comparison.} We present a quantitative evaluation of our approach in Table~\ref{tab:reds_baseline}. Compared to baseline models, our method obtains lower MSE, higher PSNR, and higher SSIM. Specifically, the proposed \OurModel~improves the current state-of-the-art algorithm by 37.4\% in MSE, 17.8\% in PSNR, and 23.3\% in SSIM.

Qualitatively, we compare \OurModel~with baseline models in Figure~\ref{fig:reds_1} and Figure~\ref{fig:real_1}. Visually, our results are not only sharper but also less noisy than the EDI~\cite{pan2020high} reconstruction. The blurriness of EDI can be explained by their use of total variation as an image quality prior, which penalizes spatial edge features. By contrast, our method implicitly learns a deep prior from the sharp images in the training dataset. In terms of the noises, EDI assumes the event threshold can precisely characterize all intensity changes. As discussed in Section~\ref{Sec:Preliminary:Challenges}, this assumption is inaccurate and susceptible to artifacts. Compared to eSL-Net~\cite{wang2020event}, our model does not over exaggerate visual features. As shown in the second row of Figure~\ref{fig:reds_1}, the eSL-Net output contains a distorted patch in the bottom-right corner, an overemphasis of the floor tile. The excessive amount of noise makes eSL-Net unfavorable in the quantitative evaluation, even though subjectively, it reconstructs sharper frames than EDI. We also invite readers to watch our supplementary animations. These animations demonstrate that our method generates temporally smooth reconstruction, while the baseline approaches suffer from trajectory discontinuity to varying degrees. 

\noindent\textbf{Ablation Study.} The input to our model has two components: the blurry frame and the events. Prior works have approached the video reconstruction problem using both the blurry frame alone~\cite{jin2018learning, purohit2019bringing} and the events alone~\cite{Rebecq19cvpr, Rebecq19pami, 9337171}. We begin our ablation study by examining the benefit of using a dual-stream input. As shown the first three rows in Table~\ref{tab:reds_ablation}, removing either input from the pipeline leads to noticeable performance degradation. For example, the MSE of the dual-input model is 6.7\% lower than the event-only model and 30.6\% lower than the frame-only model. This suggests that conventional frames and events are complementary to each other, and our proposed \OurModel~is able to take collaborative advantage of the combined information.

To examine if the polynomial video representation is indeed superior to the traditional frame-based representation, we use the same U-Net architecture to regress the $d=14$ ground-truth sharp frames directly. The comparison between the third and fourth rows in Table~\ref{tab:reds_ablation} shows that our proposed polynomial representation outperforms the frame-based baseline by 8.1\% in MSE. This result demonstrates power of explicit derivative modeling in event data.

Finally, we examine the effectiveness of refinement. Between the third and the fifth row in Table~\ref{tab:reds_ablation}, we observe that the refinement module improves the initialization by 8.8\% in MSE. The supplementary material includes an animated comparison between the initialization and refinement frames. While the initialization results successfully recovers the motion trajectory, the visual features are occasionally not sharp enough. The refinement algorithm is able to sharpen the initial results by encouraging high-quality visual features to propagate among consecutive frames.

%% file: 06-Limitations.tex
\section{Limitations}
\label{Sec:Limitations}
We point out that standard image quality metrics have a negative bias towards approaches that generate sharp but noisy results, such as eSL-Net~\cite{wang2020event}. Ideally, we would like to separate the visual signal and the noise and measure them independently. Practically, we have to resort to distance-based metrics without the disentanglement. Readers are encouraged to compare our approach with the baselines visually while using the quantitative evaluation as a reference. 

%% file: 07-Conclusion.tex
\section{Conclusion}
\label{Sec:Conclusion}
This paper introduces \OurModel, a novel event-enhanced deblurring approach that represents the intensity signal as a continuous parametric function. Experiments show that \OurModel~outperforms current state-of-the-art models in reconstruction quality. In the future, we plan to extend the approach and integrate spatial continuity into the formulation. Another possible direction is to explore probabilistic inference since each blurry image corresponds to more than one possible realistic motion trajectory.

%% file: 08-Acknowledgement.tex
\section{Acknowledgement}
\label{Sec:Acknowledgement}
We would like to sincerely appreciate Jiajia Yu (RPI), Jiacheng Chen (SFU), Zhenpei Yang (UT Austin), Bo Sun (UT Austin), Siming Yan (UT Austin), Haitao Yang (UT Austin), Jiaru Song (UT Austin), Yan He (Rice), and Ao Li (Yext) for their constructive criticism and valuable suggestions. Additionally, Qixing Huang acknowledges the support from NSF Career IIS-2047677 and NSF HDR TRIPODS1934932. Chandrajit Bajaj acknowledges the support from the NIH (DK129979), in part from the Peter O’Donnell Foundation, and in part from a grant from the Army Research Office accomplished under Cooperative Agreement Number W911NF-19-2-0333.

%% file: arXiv.bbl
\begin{thebibliography}{10}\itemsep=-1pt

\bibitem{babacan2012bayesian}
S~Derin Babacan, Rafael Molina, Minh~N Do, and Aggelos~K Katsaggelos.
\newblock Bayesian blind deconvolution with general sparse image priors.
\newblock In {\em European conference on computer vision}, pages 341--355.
  Springer, 2012.

\bibitem{6889103}
Christian Brandli, Raphael Berner, Minhao Yang, Shih-Chii Liu, and Tobi
  Delbruck.
\newblock A 240 × 180 130 db 3 µs latency global shutter spatiotemporal
  vision sensor.
\newblock {\em IEEE Journal of Solid-State Circuits}, 49(10):2333--2341, 2014.

\bibitem{caballero2017real}
Jose Caballero, Christian Ledig, Andrew Aitken, Alejandro Acosta, Johannes
  Totz, Zehan Wang, and Wenzhe Shi.
\newblock Real-time video super-resolution with spatio-temporal networks and
  motion compensation.
\newblock In {\em Proceedings of the IEEE Conference on Computer Vision and
  Pattern Recognition}, pages 4778--4787, 2017.

\bibitem{9337171}
Pablo Rodrigo~Gantier Cadena, Yeqiang Qian, Chunxiang Wang, and Ming Yang.
\newblock Spade-e2vid: Spatially-adaptive denormalization for event-based video
  reconstruction.
\newblock {\em IEEE Transactions on Image Processing}, 30:2488--2500, 2021.

\bibitem{fergus2006removing}
Rob Fergus, Barun Singh, Aaron Hertzmann, Sam~T Roweis, and William~T Freeman.
\newblock Removing camera shake from a single photograph.
\newblock In {\em ACM SIGGRAPH 2006 Papers}, pages 787--794, 2006.

\bibitem{fish1995blind}
DA Fish, AM Brinicombe, ER Pike, and JG Walker.
\newblock Blind deconvolution by means of the richardson--lucy algorithm.
\newblock {\em JOSA A}, 12(1):58--65, 1995.

\bibitem{han2021evintsr}
Jin Han, Yixin Yang, Chu Zhou, Chao Xu, and Boxin Shi.
\newblock Evintsr-net: Event guided multiple latent frames reconstruction and
  super-resolution.
\newblock In {\em Proceedings of the IEEE/CVF International Conference on
  Computer Vision}, pages 4882--4891, 2021.

\bibitem{jiang2020learning}
Zhe Jiang, Yu Zhang, Dongqing Zou, Jimmy Ren, Jiancheng Lv, and Yebin Liu.
\newblock Learning event-based motion deblurring.
\newblock In {\em Proceedings of the IEEE/CVF Conference on Computer Vision and
  Pattern Recognition}, pages 3320--3329, 2020.

\bibitem{jin2018learning}
Meiguang Jin, Givi Meishvili, and Paolo Favaro.
\newblock Learning to extract a video sequence from a single motion-blurred
  image.
\newblock In {\em Proceedings of the IEEE Conference on Computer Vision and
  Pattern Recognition}, pages 6334--6342, 2018.

\bibitem{joshi2009image}
Neel Joshi, C~Lawrence Zitnick, Richard Szeliski, and David~J Kriegman.
\newblock Image deblurring and denoising using color priors.
\newblock In {\em 2009 IEEE Conference on Computer Vision and Pattern
  Recognition}, pages 1550--1557. IEEE, 2009.

\bibitem{713236}
Sang~Ku Kim, Sang~Rae Park, and Joon~Ki Paik.
\newblock Simultaneous out-of-focus blur estimation and restoration for digital
  auto-focusing system.
\newblock {\em IEEE Transactions on Consumer Electronics}, 44(3):1071--1075,
  1998.

\bibitem{kingma2014adam}
Diederik~P Kingma and Jimmy Ba.
\newblock Adam: A method for stochastic optimization.
\newblock {\em arXiv preprint arXiv:1412.6980}, 2014.

\bibitem{krishnan2009fast}
Dilip Krishnan and Rob Fergus.
\newblock Fast image deconvolution using hyper-laplacian priors.
\newblock {\em Advances in neural information processing systems},
  22:1033--1041, 2009.

\bibitem{kupyn2018deblurgan}
Orest Kupyn, Volodymyr Budzan, Mykola Mykhailych, Dmytro Mishkin, and
  Ji{\v{r}}{\'\i} Matas.
\newblock Deblurgan: Blind motion deblurring using conditional adversarial
  networks.
\newblock In {\em Proceedings of the IEEE conference on computer vision and
  pattern recognition}, pages 8183--8192, 2018.

\bibitem{kupyn2019deblurgan}
Orest Kupyn, Tetiana Martyniuk, Junru Wu, and Zhangyang Wang.
\newblock Deblurgan-v2: Deblurring (orders-of-magnitude) faster and better.
\newblock In {\em Proceedings of the IEEE/CVF International Conference on
  Computer Vision}, pages 8878--8887, 2019.

\bibitem{lacoste2019quantifying}
Alexandre Lacoste, Alexandra Luccioni, Victor Schmidt, and Thomas Dandres.
\newblock Quantifying the carbon emissions of machine learning.
\newblock {\em arXiv preprint arXiv:1910.09700}, 2019.

\bibitem{levin2007image}
Anat Levin, Rob Fergus, Fr{\'e}do Durand, and William~T Freeman.
\newblock Image and depth from a conventional camera with a coded aperture.
\newblock {\em ACM transactions on graphics (TOG)}, 26(3):70--es, 2007.

\bibitem{li2019fast}
Sheng Li, Fengxiang He, Bo Du, Lefei Zhang, Yonghao Xu, and Dacheng Tao.
\newblock Fast spatio-temporal residual network for video super-resolution.
\newblock In {\em Proceedings of the IEEE/CVF Conference on Computer Vision and
  Pattern Recognition}, pages 10522--10531, 2019.

\bibitem{1696265}
P. Lichtsteiner, C. Posch, and T. Delbruck.
\newblock A 128 x 128 120db 30mw asynchronous vision sensor that responds to
  relative intensity change.
\newblock In {\em 2006 IEEE International Solid State Circuits Conference -
  Digest of Technical Papers}, pages 2060--2069, 2006.

\bibitem{4444573}
Patrick Lichtsteiner, Christoph Posch, and Tobi Delbruck.
\newblock A 128$\times$ 128 120 db 15 $\mu$s latency asynchronous temporal
  contrast vision sensor.
\newblock {\em IEEE Journal of Solid-State Circuits}, 43(2):566--576, 2008.

\bibitem{lin2020learning}
Songnan Lin, Jiawei Zhang, Jinshan Pan, Zhe Jiang, Dongqing Zou, Yongtian Wang,
  Jing Chen, and Jimmy Ren.
\newblock Learning event-driven video deblurring and interpolation.
\newblock In {\em Computer Vision--ECCV 2020: 16th European Conference,
  Glasgow, UK, August 23--28, 2020, Proceedings, Part VIII 16}, pages 695--710.
  Springer, 2020.

\bibitem{Nah_2019_CVPR_Workshops_REDS}
Seungjun Nah, Sungyong Baik, Seokil Hong, Gyeongsik Moon, Sanghyun Son, Radu
  Timofte, and Kyoung~Mu Lee.
\newblock Ntire 2019 challenge on video deblurring and super-resolution:
  Dataset and study.
\newblock In {\em CVPR Workshops}, June 2019.

\bibitem{niklaus2017video}
Simon Niklaus, Long Mai, and Feng Liu.
\newblock Video frame interpolation via adaptive separable convolution.
\newblock In {\em Proceedings of the IEEE International Conference on Computer
  Vision}, pages 261--270, 2017.

\bibitem{Paikin_2021_CVPR}
Genady Paikin, Yotam Ater, Roy Shaul, and Evgeny Soloveichik.
\newblock Efi-net: Video frame interpolation from fusion of events and frames.
\newblock In {\em Proceedings of the IEEE/CVF Conference on Computer Vision and
  Pattern Recognition (CVPR) Workshops}, pages 1291--1301, June 2021.

\bibitem{pan2020high}
Liyuan Pan, Richard Hartley, Cedric Scheerlinck, Miaomiao Liu, Xin Yu, and
  Yuchao Dai.
\newblock High frame rate video reconstruction based on an event camera.
\newblock {\em IEEE Transactions on Pattern Analysis and Machine Intelligence},
  2020.

\bibitem{pan2019bringing}
Liyuan Pan, Cedric Scheerlinck, Xin Yu, Richard Hartley, Miaomiao Liu, and
  Yuchao Dai.
\newblock Bringing a blurry frame alive at high frame-rate with an event
  camera.
\newblock In {\em Proceedings of the IEEE Conference on Computer Vision and
  Pattern Recognition}, pages 6820--6829, 2019.

\bibitem{park2020vid}
Sunghyun Park, Kangyeol Kim, Junsoo Lee, Jaegul Choo, Joonseok Lee, Sookyung
  Kim, and Edward Choi.
\newblock Vid-ode: Continuous-time video generation with neural ordinary
  differential equation.
\newblock {\em arXiv preprint arXiv:2010.08188}, 2020.

\bibitem{6909768}
Daniele Perrone and Paolo Favaro.
\newblock Total variation blind deconvolution: The devil is in the details.
\newblock In {\em 2014 IEEE Conference on Computer Vision and Pattern
  Recognition}, pages 2909--2916, 2014.

\bibitem{purohit2019bringing}
Kuldeep Purohit, Anshul Shah, and AN Rajagopalan.
\newblock Bringing alive blurred moments.
\newblock In {\em Proceedings of the IEEE/CVF Conference on Computer Vision and
  Pattern Recognition}, pages 6830--6839, 2019.

\bibitem{Rebecq18corl}
Henri Rebecq, Daniel Gehrig, and Davide Scaramuzza.
\newblock {ESIM}: an open event camera simulator.
\newblock {\em Conf. on Robotics Learning (CoRL)}, Oct. 2018.

\bibitem{Rebecq19cvpr}
Henri Rebecq, Ren{\'{e}} Ranftl, Vladlen Koltun, and Davide Scaramuzza.
\newblock Events-to-video: Bringing modern computer vision to event cameras.
\newblock {\em {IEEE} Conf. Comput. Vis. Pattern Recog. (CVPR)}, 2019.

\bibitem{Rebecq19pami}
Henri Rebecq, Ren{\'{e}} Ranftl, Vladlen Koltun, and Davide Scaramuzza.
\newblock High speed and high dynamic range video with an event camera.
\newblock {\em {IEEE} Trans. Pattern Anal. Mach. Intell. (T-PAMI)}, 2019.

\bibitem{richardson1972bayesian}
William~Hadley Richardson.
\newblock Bayesian-based iterative method of image restoration.
\newblock {\em JoSA}, 62(1):55--59, 1972.

\bibitem{ronneberger2015u}
Olaf Ronneberger, Philipp Fischer, and Thomas Brox.
\newblock U-net: Convolutional networks for biomedical image segmentation.
\newblock In {\em International Conference on Medical image computing and
  computer-assisted intervention}, pages 234--241. Springer, 2015.

\bibitem{shan2008high}
Qi Shan, Jiaya Jia, and Aseem Agarwala.
\newblock High-quality motion deblurring from a single image.
\newblock {\em Acm transactions on graphics (tog)}, 27(3):1--10, 2008.

\bibitem{shang2021bringing}
Wei Shang, Dongwei Ren, Dongqing Zou, Jimmy~S Ren, Ping Luo, and Wangmeng Zuo.
\newblock Bringing events into video deblurring with non-consecutively blurry
  frames.
\newblock In {\em Proceedings of the IEEE/CVF International Conference on
  Computer Vision}, pages 4531--4540, 2021.

\bibitem{tian2018tdan}
Yapeng Tian, Yulun Zhang, Yun Fu, and Chenliang Xu.
\newblock Tdan: Temporally deformable alignment network for video
  super-resolution.
\newblock {\em arXiv preprint arXiv:1812.02898}, 2018.

\bibitem{tian2020tdan}
Yapeng Tian, Yulun Zhang, Yun Fu, and Chenliang Xu.
\newblock Tdan: Temporally-deformable alignment network for video
  super-resolution.
\newblock In {\em The IEEE Conference on Computer Vision and Pattern
  Recognition (CVPR)}, June 2020.

\bibitem{tulyakov2021time}
Stepan Tulyakov, Daniel Gehrig, Stamatios Georgoulis, Julius Erbach, Mathias
  Gehrig, Yuanyou Li, and Davide Scaramuzza.
\newblock Time lens: Event-based video frame interpolation.
\newblock In {\em Proceedings of the IEEE/CVF Conference on Computer Vision and
  Pattern Recognition}, pages 16155--16164, 2021.

\bibitem{wang2020event}
Bishan Wang, Jingwei He, Lei Yu, Gui-Song Xia, and Wen Yang.
\newblock Event enhanced high-quality image recovery.
\newblock In {\em European Conference on Computer Vision}. Springer, 2020.

\bibitem{xu2021motion}
Fang Xu, Lei Yu, Bishan Wang, Wen Yang, Gui-Song Xia, Xu Jia, Zhendong Qiao,
  and Jianzhuang Liu.
\newblock Motion deblurring with real events.
\newblock In {\em Proceedings of the IEEE/CVF International Conference on
  Computer Vision}, pages 2583--2592, 2021.

\bibitem{xu2010two}
Li Xu and Jiaya Jia.
\newblock Two-phase kernel estimation for robust motion deblurring.
\newblock In {\em European conference on computer vision}, pages 157--170.
  Springer, 2010.

\bibitem{xu2013unnatural}
Li Xu, Shicheng Zheng, and Jiaya Jia.
\newblock Unnatural l0 sparse representation for natural image deblurring.
\newblock In {\em Proceedings of the IEEE conference on computer vision and
  pattern recognition}, pages 1107--1114, 2013.

\bibitem{zhang2021fine}
Limeng Zhang, Hongguang Zhang, Chenyang Zhu, Shasha Guo, Jihua Chen, and Lei
  Wang.
\newblock Fine-grained video deblurring with event camera.
\newblock In {\em International Conference on Multimedia Modeling}, pages
  352--364. Springer, 2021.

\end{thebibliography}
